\documentclass[
 aip,
 amsmath,amssymb,
 reprint%
]{revtex4-2}

\usepackage{graphicx}
\usepackage{dcolumn}
\usepackage{bm}
\usepackage[utf8x]{inputenc} 
\usepackage{todonotes}
\usepackage{svg}
\usepackage{siunitx}
\usepackage{upgreek}

\usepackage[mathlines]{lineno}

\begin{document}

\preprint{AAPM/123-QED}

\title{Single-photon detection in the mid-infrared up to 10 micron wavelength using tungsten silicide superconducting nanowire detectors}

\author{V. B. Verma}
\email{verma@nist.gov}
\affiliation{National Institute of Standards and Technology, Boulder, CO, USA.}
\author{B. Korzh}%
\email{bkorzh@jpl.nasa.gov}
\affiliation{Jet Propulsion Laboratory, California Institute of Technology, 4800 Oak Grove Dr., Pasadena, CA, USA}
\author{A. B. Walter}
\affiliation{Jet Propulsion Laboratory, California Institute of Technology, 4800 Oak Grove Dr., Pasadena, CA, USA}
\author{A. E. Lita}
\affiliation{National Institute of Standards and Technology, Boulder, CO, USA.}
\author{R. M. Briggs}
\affiliation{Jet Propulsion Laboratory, California Institute of Technology, 4800 Oak Grove Dr., Pasadena, CA, USA}
\author{M. Colangelo}
 \affiliation{Department of Electrical Engineering and Computer Science,
Massachusetts Institute of Technology, Cambridge, MA, USA.}%
\author{Y. Zhai}
\affiliation{National Institute of Standards and Technology, Boulder, CO, USA.}
\author{E. E. Wollman}
\affiliation{Jet Propulsion Laboratory, California Institute of Technology, 4800 Oak Grove Dr., Pasadena, CA, USA}
\author{A. D. Beyer}
\affiliation{Jet Propulsion Laboratory, California Institute of Technology, 4800 Oak Grove Dr., Pasadena, CA, USA}
\author{J. P. Allmaras}
\affiliation{Jet Propulsion Laboratory, California Institute of Technology, 4800 Oak Grove Dr., Pasadena, CA, USA}
\author{B. Bumble}
\affiliation{Jet Propulsion Laboratory, California Institute of Technology, 4800 Oak Grove Dr., Pasadena, CA, USA}
\author{H. Vora}
\affiliation{National Institute of Standards and Technology, Boulder, CO, USA.}
\author{D. Zhu}
 \affiliation{Department of Electrical Engineering and Computer Science,
Massachusetts Institute of Technology, Cambridge, MA, USA.}%
\author{E. Schmidt}
\affiliation{Jet Propulsion Laboratory, California Institute of Technology, 4800 Oak Grove Dr., Pasadena, CA, USA}
\author{K. K. Berggren}
 \affiliation{Department of Electrical Engineering and Computer Science,
Massachusetts Institute of Technology, Cambridge, MA, USA.}%
\author{R. P. Mirin}
\affiliation{National Institute of Standards and Technology, Boulder, CO, USA.}
\author{S. W. Nam}
\affiliation{National Institute of Standards and Technology, Boulder, CO, USA.}
\author{M. D. Shaw}
\affiliation{Jet Propulsion Laboratory, California Institute of Technology, 4800 Oak Grove Dr., Pasadena, CA, USA}

\date{\today}

\begin{abstract}
We developed superconducting nanowire single-photon detectors (SNSPDs) based on tungsten silicide (WSi) that show saturated internal detection efficiency up to a wavelength of 10~\si{\upmu m}. These detectors are promising for applications in the mid-infrared requiring ultra-high gain stability, low dark counts, and high efficiency such as chemical sensing, LIDAR, dark matter searches and exoplanet spectroscopy. 
\end{abstract}

\maketitle

Efficient single-photon counting, with a detection efficiency greater than 50\% has, to date, been achieved only at wavelengths shorter than 2~\si{\upmu m}~\cite{Eisaman2011}. Extension of such performance to the mid-infrared has potential for new applications in astronomy~\cite{OST2019} as well as LIDAR~\cite{Taylor2019}, dark matter searches~\cite{Hochberg2019} and the fundamental study of fast molecular dynamics and chemistry~\cite{Chen2017, Chen2018, Chen2019, Lau2020}. Interest in the mid-infrared stems primarily from the presence of numerous absorption signatures for molecules such as water vapor, carbon dioxide, oxygen and ozone, methane, and nitrous oxide. These molecules are not only important for understanding of the evolution of our own planet, but may also be signatures of life or indicate the potential habitability of planets outside of our solar system. The spectroscopy of exoplanet atmospheres is a prime example of an application requiring single-photon-counting detectors with ultra-stable gain and high efficiency due to the photon-starved nature of the measurements~\cite{Deming_2009, Greene_2016}.  

Here we report on superconducting nanowire single photon detectors (SNSPDs)  based on tungsten silicide (WSi) with saturated internal detection efficiency up to a wavelength of 10~\si{\upmu m}. Internal detection efficiency is defined as the probability that a photon absorbed in the nanowire generates an electrical output pulse from the detector. Prior work in the near-infrared has shown that achieving saturated internal detection efficiency is a critical step in making high efficiency devices~\cite{Marsili2013, Reddy2020}. Our results indicate that it may be possible to achieve high system detection efficiency in the mid-infrared with WSi SNSPDs by optimizing the absorption in the nanowires and the coupling to the active area of the detectors, as has been done at near-infrared wavelengths.

Previous approaches to achieving sensitivity in the mid-infrared were based on SNSPDs fabricated from niobium nitride (NbN) thin films. NbN has historically been the material of choice for SNSPDs due to its relatively high superconducting transition temperature (8 \si{Kelvin} to 12 \si{Kelvin}) allowing operation at 4 \si{Kelvin}, and its fast thermal recovery time allowing operation at high count rates~\cite{Vetter2016}. In order to improve the energy sensitivity of the nanowire detectors and extend the photo-response to longer wavelengths, there are several possible approaches. One is to decrease the cross-sectional area of the nanowire, making the film thinner or the nanowire narrower. This results in a higher probability of a hotspot (non-superconducting domain) being generated since the energy per unit area is larger, and thermal conductivity along the length of the nanowire is smaller. Using this approach, saturated internal detection efficiency has been demonstrated with NbN SNSPDs up to a wavelength of approximately 3 \si{\upmu m} using ultra-narrow 30 nm-wide nanowires~\cite{Marsili2012}. In another report, single-photon sensitivity was achieved at a wavelength of 10.6~\si{\upmu m} using 40 nm-wide NbN nanowires, although the internal detection efficiency was not saturated\cite{Korneev2012}.

The second approach to improving energy sensitivity is to engineer the material to reduce the free carrier density. Reducing the free carrier density (increasing resistivity) of the material results in the deposited energy being divided amongst fewer quasiparticles, thus increasing the effective temperature of each quasiparticle. Simultaneously, the thermal impedance along the length of the nanowire is increased, which more effectively localizes the deposited energy along its length. This approach is preferable to reducing the width and thickness of the nanowires from a fabrication and yield perspective, as detector yield begins to degrade significantly at widths below about 80 nm~\cite{Frasca2019}. Below this width, edge roughness and constrictions in the width of the nanowire due to fabrication imperfections begin to dominate and significantly degrade the device performance. The constrictions result in a suppressed switching current, the current at which the critical current density in the wire is exceeded and the superconductor switches to the normal (non-superconducting) state. 

Yet another approach to improving energy sensitivity is reducing the superconducting gap energy. This results in a larger number of broken Cooper pairs (quasiparticles) for a given amount of deposited energy in the superconductor~\cite{Allmaras2020thesis}. However, reducing the superconducting gap energy implies a lower superconducting transition temperature ($T_\mathrm{c}$) and thus a lower operating temperature for the detectors. This is generally undesirable as it results in an increase in the complexity and cost of the cryogenics. 
	
\begin{figure}[htbp]
\centering
\includegraphics[width=\linewidth]{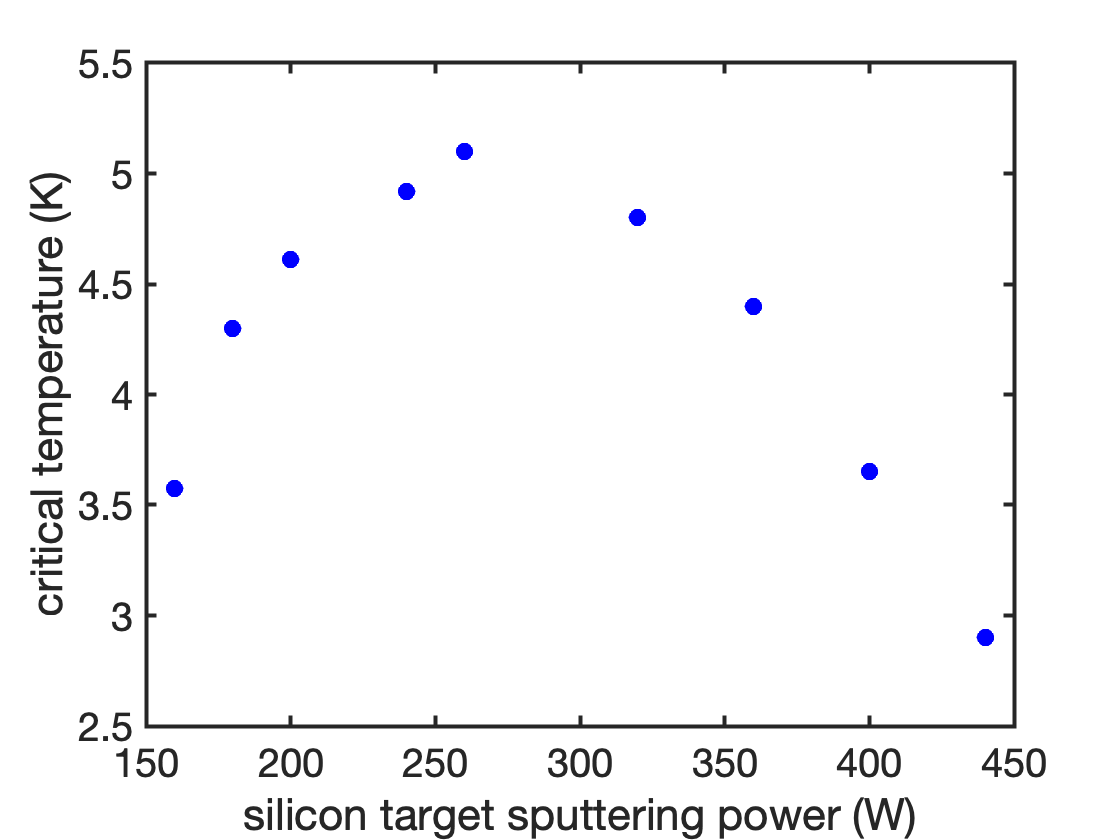}
\caption{Superconducting transition temperature ($T_\mathrm{c}$) as a function of sputtering power on the silicon target during co-sputtering of the WSi film. The W sputtering power is fixed at 100~W. Measurements were performed on bulk films with thickness greater than or equal to 50~nm.}
\label{TcVsPower}
\end{figure}

As outlined above, the most promising approach to lowering the energy threshold is to increase the resistivity of the superconducting film, which is typically achieved through variation of the stoichiometry. For amorphous WSi, the $T_\mathrm{c}$ is not strongly dependent on stoichiometry as shown in Fig.~\ref{TcVsPower}, which allows a significant variation in resistivity while maintaining a fixed operating temperature of the system. In addition, the films remain amorphous over a wide range of sputtering parameters and stoichiometries~\cite{Kondo1992}, as verified by x-ray diffraction measurements. 

\begin{figure}[htbp]
\centering
\includegraphics[width=\linewidth]{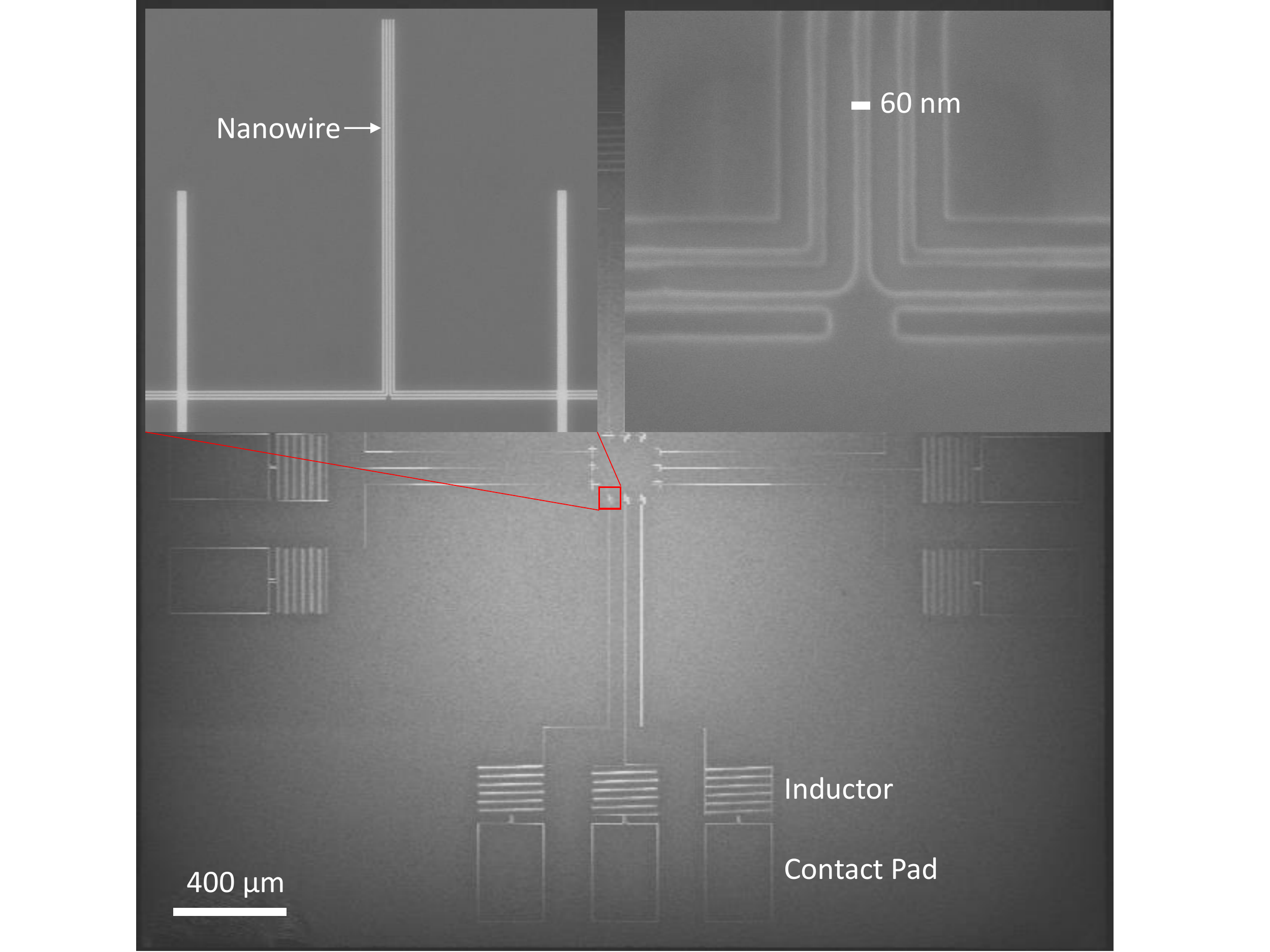}
\caption{Scanning electron micrographs of the chip layout.}
\label{SEM}
\end{figure}

We fabricated detectors from high-resistivity WSi films using a Si-rich stoichiometry.  The films are co-sputtered from separate W and Si targets, allowing the composition of the films to be tuned by adjusting the relative sputtering powers. Typical sputtering powers for the W and Si targets are 100 and 180~W, respectively, which in our deposition system results in a silicon content of ~ 15\%. We increased the Si sputtering power from 180 to 260~W, while maintaining the W sputtering power at 100~W. This resulted in a silicon composition of 35±7\% as estimated by secondary ion mass spectroscopy (SIMS). The film thickness was 3.2~nm, and the superconducting transition temperature was 3.1~K. Note that the $T_\mathrm{c}$ of the thin film is significantly lower than that shown in Fig.~\ref{TcVsPower} due to the decrease of $T_c$ with film thickness.
Each detector is a single 10~\si{\upmu m}-long nanowire instead of a large-area meander, in order to reduce the probability of fabrication defects or constrictions along the length of the wire for the relatively narrow widths investigated ranging from 50 to 80~nm. A 2~\si{\upmu H} inductance consisting of a 10~\si{\upmu m}-wide meandering wire was patterned into the WSi film in series with the nanowire to prevent latching to the non-superconducting state~\cite{Annunziata2010}. Scanning electron micrographs of the chip layout are shown in Fig.~\ref{SEM}. The small active area of the device prevents it from achieving a high coupling efficiency in mid-infrared applications, however, it serves as a well controlled experiment to study the optimum nanowire material and geometry to achieve the maximum intrinsic detection efficiency. 

\begin{figure}[htbp]
\centering
\fbox{\includegraphics[width=\linewidth]{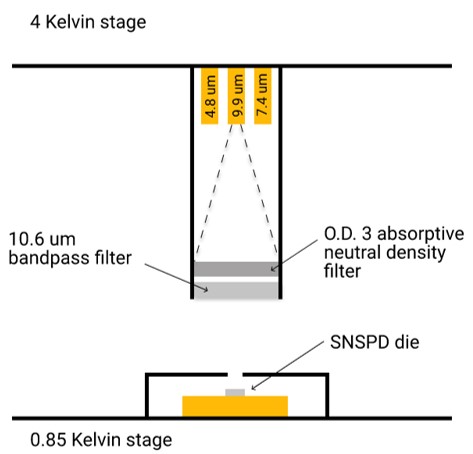}}
\caption{Quantum cascade laser (QCL) setup inside the cryostat, flood illuminating the sample. Typically the 10.6~\si{\upmu m} bandpass filter was omitted, in order to enable the use of all three lasers on the same cooldown. But it was used to verify the response at 9.9~\si{\upmu m} with and without it.  
}
\label{QCLSetup}
\end{figure}

\begin{figure}[htbp]
\centering
\includegraphics[width=\linewidth]{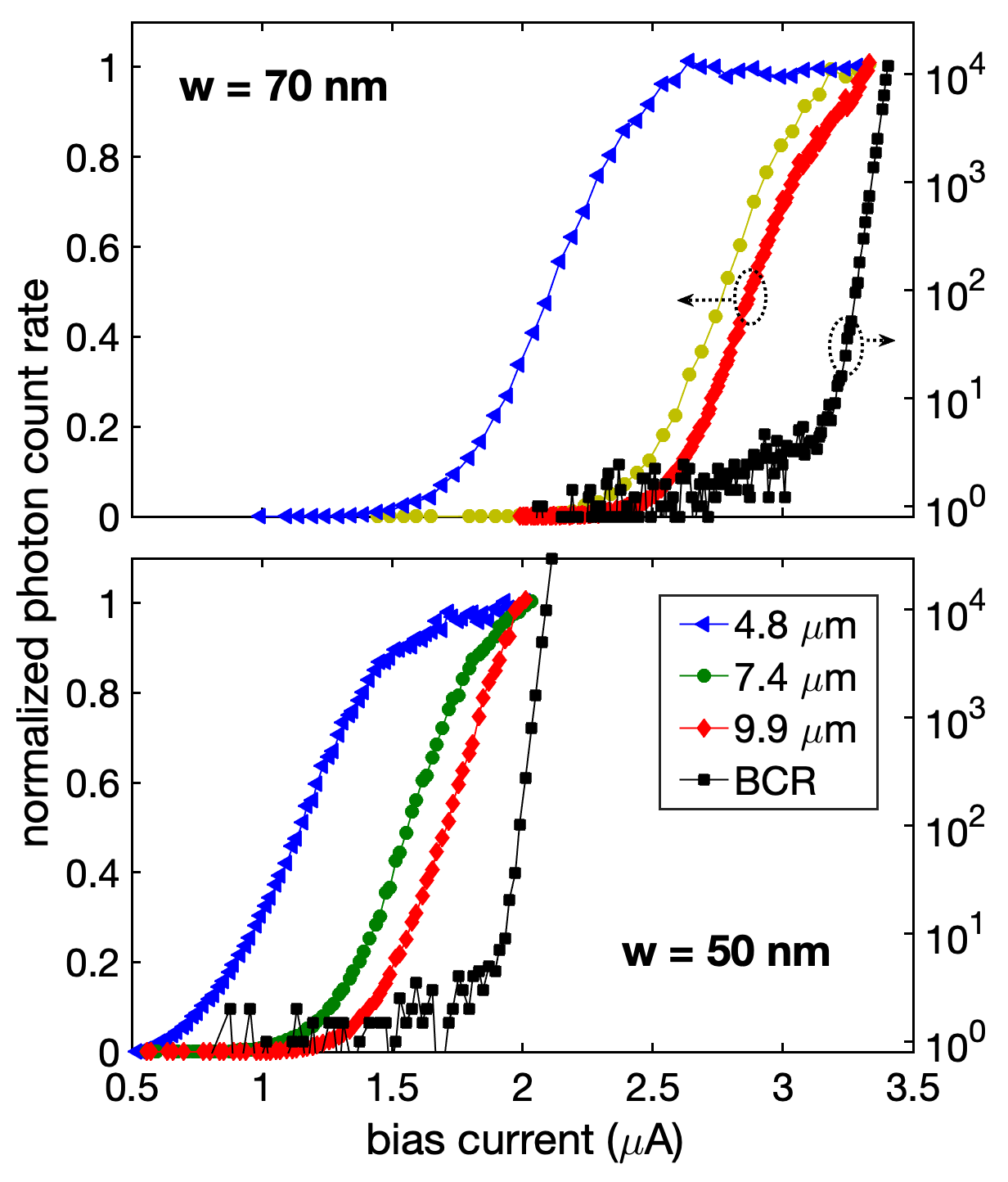}
\caption{Normalized photoresponse count rate vs. bias current curves for SNSPDs fabricated from a WSi film with a higher silicon content (sputtering power on silicon target increased from 180 to 260~W). PCR curves are shown for two different nanowire widths (50 and 70~nm) and measurements were obtained at three wavelengths (4.8, 7.4, and 9.9~\si{\upmu m}). Black squares correspond to measurement of the background count rates.}

\label{PCR1}
\end{figure}

The SNSPDs were measured at a temperature of 0.9~K, and flood illuminated using various quantum cascade lasers (QCLs) also mounted inside the cryostat, at the 4~Kelvin stage. Figure~\ref{QCLSetup} shows the measurement setup. The three QCLs (4.8, 7.4, and 9.9~\si{\upmu m}) are mounted side by side, such that it is possible to study the response to different wavelengths during the same cooldown of the cryostat. The QCLs are pulsed with 260~\si{\upmu s} pulses and a 4\% duty cycle. To collect the photon response curves, a 200~\si{\upmu s} gate is applied on a counter, synchronized with the middle of the laser pulses. The background rate (BCR) is collected by synchronizing the gate with the period when the laser is turned off. The signal from the SNSPD is amplified with a cryogenic amplifier operating at the 4~kelvin stage, with a gain of 45~dB, bandwidth of 1.5~GHz and noise temperature of less than 6~K. An additional room temperature amplifier with 25~dB of gain and 500~MHz bandwidth was used, followed by a 120~MHz low-pass filter. This readout scheme enabled pulses to be readout while biased with currents as low as 200~nA. Figure~\ref{PCR1} shows normalized photon count rate (PCR) vs. bias current curves for SNSPDs fabricated with a film deposited with a sputtering power on the silicon target of 260~W. PCR curves are shown for two different nanowire widths (50 \si{nm} and 70~nm), and measurements were obtained at three wavelengths (4.8, 7.4, and 9.9~\si{\upmu m}). As shown in the figure, the 70 nm-wide nanowire showed the onset of a saturation plateau even at a wavelength as long as 9.9~\si{\upmu m}. The absence of this feature for narrower wires is likely due to the presence of a fabrication defect such as a constriction in the width of the nanowire. 

\begin{figure}[htbp]
\centering
\includegraphics[width=\linewidth]{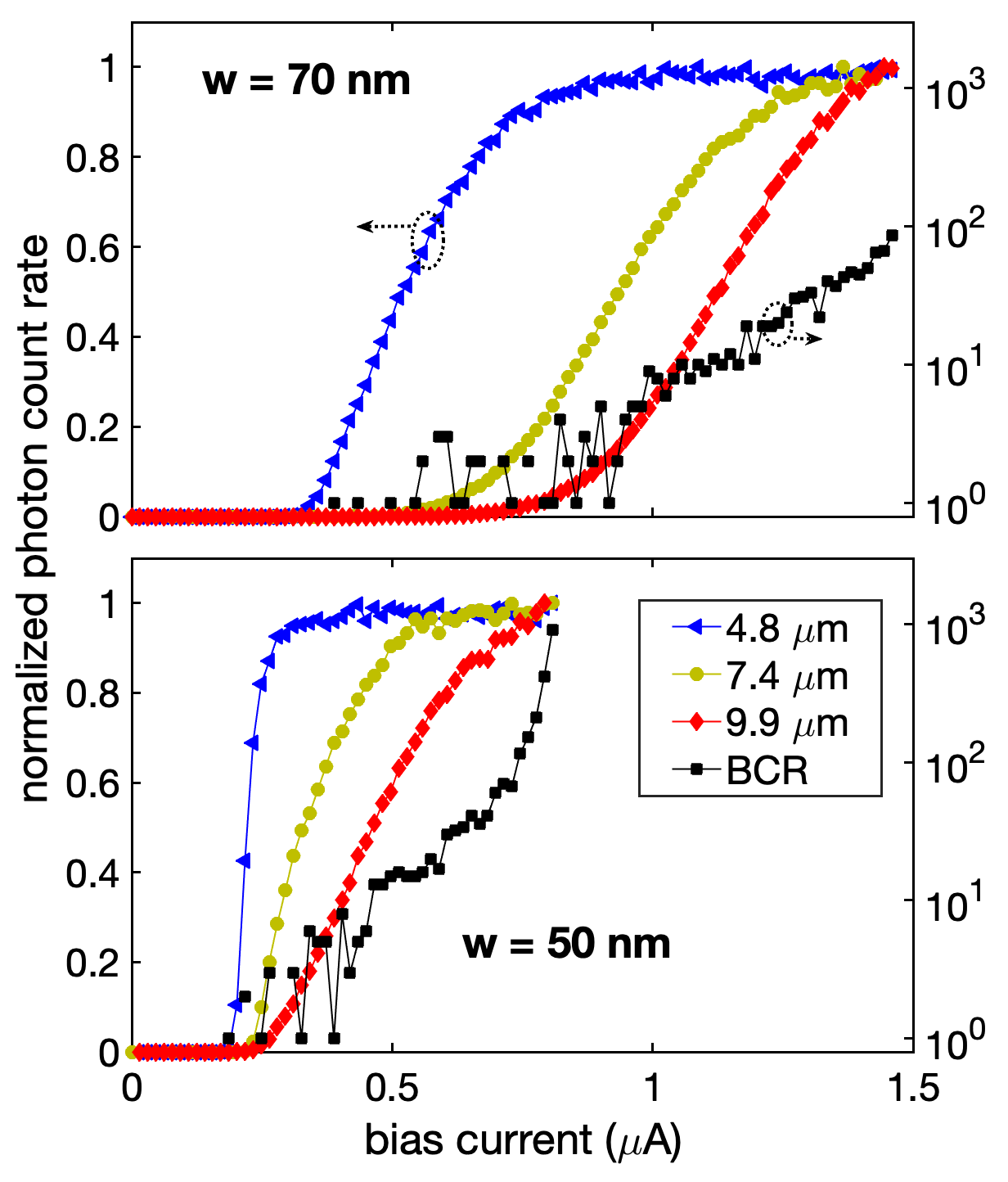}
\caption{Normalized photoresponse count rate vs. bias current curves for SNSPDs fabricated from a WSi film with a further increase of the silicon content (sputtering power on silicon target increased from 260~W to 320~W) compared to the devices in Fig.~\ref{PCR1}. Two different nanowire widths (50 and 70~nm) are presented and measurements were obtained at three wavelengths (4.8, 7.4, and 9.9~\si{\upmu m}). Black squares correspond to measurement of the background count rates. The switching currents are lower, however, the relative saturation of the internal efficiency is better for all wavelengths and nanowire geometries.}
\label{PCR2}
\end{figure}

In order to further explore the effect of film stoichiometry on energy sensitivity, we fabricated a second set of devices with an even higher sputtering power on the silicon target of 320~W. Analysis by secondary ion mass spectroscopy indicated a silicon composition of 48 ± 10\%. The film thickness was determined to be 2.61~nm, slightly thinner than the previous set of devices with the lower silicon sputtering power, with a marginally lower Tc of 2.8~K compared to the previous film with a Tc of 3.1~K. Fig.~\ref{PCR2} shows the photon count rate vs. bias current curves and dark count measurements for two nanowires having widths of 70~nm and 50~nm and a length of 10~\si{\upmu m}. Both geometries show a significant improvement in saturation of the internal detection efficiency at all three wavelengths. Note that the shape of the 4.8~\si{\upmu m} PCR curve for the 50~nm wire is distorted at the lowest bias currents because the readout threshold was about 200~nA, hence any pulses created at lower currents were missed by the counter.
              
In conclusion, we have fabricated WSi SNSPDs with a saturated internal detection efficiency at mid-infrared wavelengths up to 10~\si{\upmu m}. Increasing film resistivity by tuning the film stoichiometry appears to be a promising approach to improving energy sensitivity. Demonstrating saturated internal efficiency is the first important step to obtaining high system detection efficiency by efficient design of an optical stack around the detector to enhance absorption~\cite{Reddy2020}, and optimized self-aligned fiber coupling~\cite{Miller2011} or free-space coupling~\cite{Wollman2019}. However, with both fiber-coupling and free-space coupling, filtering blackbody radiation from the environment presents a significant engineering challenge in the mid-infrared, requiring the use of cold filters or gratings. While the initial data presented here appears promising, further work is required to investigate the scalability and yield of narrow wires on the order of 50 – 80~nm over larger areas, which will be required for kilopixel and larger arrays required for applications such as exoplanet spectroscopy.

\section{Acknowledgements}
Support for this work was provided in part by the Defense Advanced Research Projects Agency, Defense Sciences Office, through the Detect program. A. B. Walter's research was supported by an appointment to the NASA Postdoctoral Program at the Jet Propulsion Laboratory, administered by Universities Space Research Association under contract with NASA. Support for this work was provided in part by the NASA ROSES-APRA program. Part of this research was performed at the Jet Propulsion Laboratory, California Institute of Technology, under contract with the National Aeronautics and Space Administration (80NM0018D0004). The authors would like to thank Pierre Echternach for technical discussions.

\bibliography{mir}

\end{document}